\documentclass[graybox,natbib,nosecnum]{svmult}
\bibpunct{(}{)}{;}{a}{}{,} 

\pdfoutput=1   

\usepackage{mathptmx}       
\usepackage{helvet}         
\usepackage{courier}        
\usepackage{type1cm}        

\usepackage{makeidx}         
\usepackage{graphicx}        
\usepackage{multicol}        
\usepackage[bottom]{footmisc}
\usepackage[normalem]{ulem}	
\usepackage{hyperref}  

\usepackage{soul}   

\newcommand*\aap{A\&A}

\newcommand*\aj{AJ}

\newcommand*\apj{ApJ}
\newcommand*\apjl{ApJ}

\newcommand*\apss{Ap\&SS}
\newcommand*\araa{ARA\&A}

\newcommand*\grl{Geophys Res Lett}

\newcommand*\icarus{Icarus}

\newcommand*\mnras{MNRAS}

\newcommand*\nat{Nature}

\newcommand*\pasj{PASJ}

\newcommand*\planss{Planet Space Sci}

\newcommand*\psj{Planetary Science Journal}

\makeindex             

\begin{document}

\title*{Planet formation theory: an overview}
\author{Philip J. Armitage}
\institute{Philip J. Armitage \at Center for Computational Astrophysics, Flatiron Institute, New York, NY 10010, USA \\
Departmant of Physics and Astronomy, Stony Brook University, Stony Brook, NY 11794, USA \\
\email{philip.armitage@stonybrook.edu}}
%
%
\maketitle

\abstract{The standard model for planet formation is a bottom-up process in which the origin of rocky and gaseous planets can be traced back to the collision of micron-sized dust grains within the gas-rich environment of protoplanetary disks. Key milestones along the way include disk formation, grain growth, planetesimal formation, core growth, gas accretion, and planetary system evolution. I provide an introductory overview of planet formation, emphasizing the main ideas and reviewing current theoretical understanding. Many of the phases of planet formation have a well-developed physical understanding, though the complexity of the problem means that few can be quantitatively modeled with complete confidence. Transformative advances in disk imaging provide the first direct information on the initial conditions for planet formation, while exoplanet data has motivated new formation models that are faster, more efficient, and lead to a more diverse set of architectures than their Solar System inspired forebears. Much remains to be learned, and I close with a personal, incomplete list, of open problems.}

\section{Introduction}
Since the work of \cite{safronov72}, planet formation has been understood to be a bottom-up process whose initial phases play out within the dust and gas of {\em protoplanetary disks}. In contemporary versions of the theory, microscopic grains collide and stick to form {\em pebbles}, which participate in collective processes that form km (and larger) scale {\em planetesimals} via self-gravity. A combination of gravitationally mediated planetesimal collisions, and aerodynamically assisted pebble accretion, drives growth of the planetesimals up to the scale of Earth masses. If disk gas remains present, it can be captured by sufficiently massive {\em cores} to form the primary atmospheres or envelopes of the Solar System's ice- and gas-giants, and of sub-Neptune and larger exoplanets. Less massive planets acquire secondary atmospheres from outgassing. Substantial orbital changes ({\em migration}) commonly occur after planets have formed, due to energy and angular momentum exchange between planets, between planets and gaseous or planetesimal disks, and between planets and their stellar hosts or binary companions.

This review provides an introduction to this standard model, and serves as an overview of the following Chapters which discuss planet formation theory in detail. I introduce the physical processes relevant for planet formation, and summarize how they fit together in our consensus understanding of how planets form. Although a plausible outline of planet formation theory has been known for decades, existing models are surely not the last word: despite transformative advances in exoplanet detection and disk imaging there remain few direct observations of planet formation processes. It is conceivable that are fundamental gaps in our current theory. Even within the established paradigm, some phases are known to be poorly understood, and I conclude by listing some open questions that I hope to see resolved soon.

\section{Disk formation and evolution}

Although limited grain growth may occur during star formation {\em prior} to disk formation \citep{kwon09,marchand23}, the vast majority of planet formation processes occur within protoplanetary disks. Disks are flattened near-equilibrium fluid structures,  whose properties are determined observationally to varying (mostly low) levels of precision. Typical stellar accretion rates scale roughly as $\dot{M} \sim 10^{-8} (M_*/M_\odot)^2 M_\odot {\rm yr}^{-1}$, with a measurement precision of $\sim 0.35 \ {\rm dex}$ and a spread at fixed stellar mass $M_*$ of more than an order of magnitude \citep{manara23}. The disk lifetime is around 3~Myr \citep{haisch01}, though the absolute ages of Young Stellar Objects are notoriously hard to determine \citep{soderblom14}. The {\em relative} ages of young clusters are more reliable. Disk masses can be estimated starting from various observables, including the dust continuum flux \citep{ansdell16,tychoniec20}, molecular line emission in commonly observed sub-mm tracers \citep[CO and its isotopologues, sometimes corrected using additional species such as N$_2$H$^+$;][]{trapman22}, HD emission \citep{mcclure16}, and (at the high mass end) rotation curves \citep{martire24,andrews24}. All of these methods have drawbacks -- they either work only for a subset of disks or have possibly large uncertainties -- such that a reliable empirical model for disk evolution remains elusive. Commonly adopted initial conditions for planet formation studies include a disk gas surface density profile that scales with radius as $\Sigma \propto r^{-1}$, an initial gas to dust ratio of $10^2$, and a mass within 100~au of $10^{-3} - 0.1 \ M_*$, but this is more a matter of art and convention than anything based on solid data. Despite its well-known shortcomings, the Minimum Mass Solar Nebula \citep{weidenschilling77b,hayashi81} also lives on in some computers.

Some important properties of protoplanetary disks follow directly from the fact that they are long-lived, and thus in near force balance. In the ``vertical" direction (perpendicular to the disk mid-plane) there is hydrostatic equilibrium between the gradient of gas pressure $P$ and the vertical component of stellar gravity,
\begin{equation}
  \frac{{\rm d}P}{{\rm d}z} = - \Omega_{\rm K}^2 \rho z,    
\end{equation}
where $\rho$ is the density, $\Omega_{\rm K}$ the Keplerian angular velocity, and we have assumed that $z \ll r$, the radial distance to the star. For a vertically isothermal disk with sound speed $c_s$, $P=\rho c_s^2$, and the gas density declines as a gaussian,
\begin{equation}
    \rho = \rho_0 \exp[-z^2 / 2 h^2],
\end{equation}
where $\rho_0$ is the mid-plane density and $h$, the {\em vertical scale height}, is,
\begin{equation}
    h \equiv \frac{c_s}{\Omega_{\rm K}}.
\end{equation}
Except in cases of unusually high accretion rates, or when close to the star, the heating of protoplanetary disks occurs primarily from the absorption and re-radiation of stellar luminosity \citep{kenyon87}. Irradiation leads to a radial temperature profile approximately given by a power-law, $T \propto r^{-3/7}$ \citep{chiang97}, and a disk that {\em flares}, such that $h/r$ increases with radius. Typically, $0.03 < h/r < 0.1$. 

In the radial direction, the force balance involves gravity, centrifugal force, and a pressure gradient,
\begin{equation}
    \Omega^2(r) r = \frac{GM_*}{r^2} + \frac{1}{\rho}\frac{{\rm d}P}{{\rm d}r}.
\end{equation}
The pressure gradient term is order $(h/r)^2$ smaller than the gravitational one, so to leading order $\Omega(r)=\Omega_{\rm K}$. The small correction -- which usually but not always leads to sub-Keplerian rotation -- is nonetheless extremely important. The dynamics of dust particles \citep{whipple72,weidenschilling77}, the formation of planetesimals \citep{youdin05}, and the interpretation of observed disk sub-structures \citep{andrews20} are all thought to hinge on this $\sim$1\% deviation from strict Keplerian motion.

\begin{figure}[t]
\includegraphics[width=\textwidth]{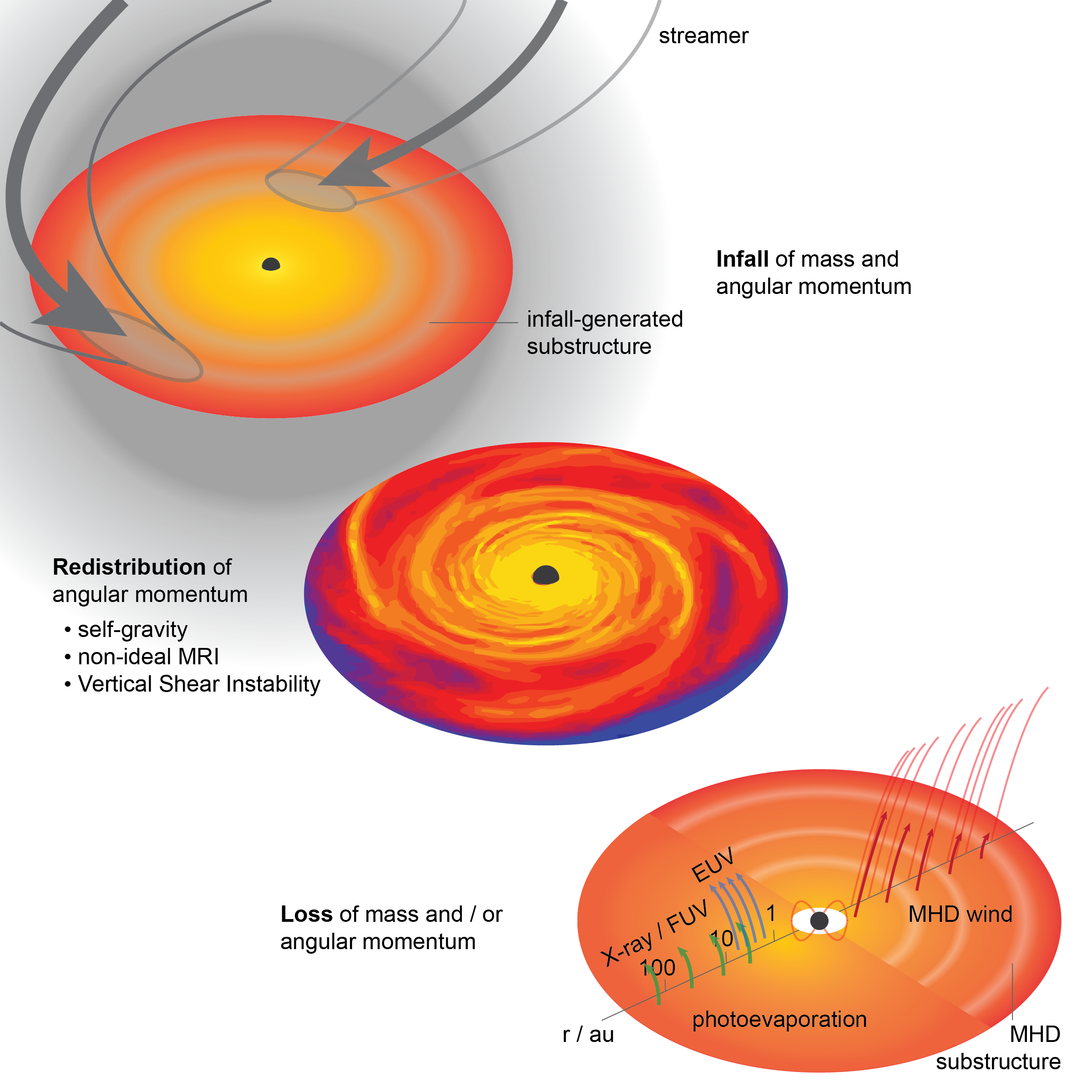}
\caption{An illustration of the main processes that lead to evolution of the gaseous component of protoplanetary disks. Observational evidence and theoretical expectations suggest a typical scenario in which the initial properties of disks are determined by infall and self-gravity, accretion is largely driven by torques from magnetohydrodynamic (MHD) winds, and dispersal is a consequence of photoevaporative and MHD outflows. Depending on the star formation environment, however, it is possible for streamers to be dynamically important at later times, or for strong external photoevaporation or stellar flybys to truncate disks early on. Turbulence, even if it is a sub-dominant process for disk evolution, remains critically important for many phases of planet formation.}
\label{Armitage_evolution_figure}
\end{figure}

In contrast to the structural questions sketched above, little can be deduced about the {\em evolution} of protoplanetary disks from elementary physical principles. (Near)-Keplerian disks have a specific angular momentum profile, $l(r) \propto r^{1/2}$, that increases with radius, so for gas to accrete angular momentum has to be lost. Broadly speaking three classes of physical processes can lead to disk evolution,
\begin{itemize}
    \item Infall of gas, which will add mass and mix angular momentum to a pre-existing disk. Infall is self-evidently important at sufficiently early times, when disks are forming and growing \citep{lin90}, but it could also be significant at late times if the density of residual gas in the disk environment is high enough \citep{throop08,winter24,padoan24}.
    \item Processes that redistribute angular momentum within the disk, while conserving disk mass (up to the fraction of gas that is physically accreted by the star). \cite{lesur23} review the long laundry list of possible processes, most of which are associated with linear instabilities of disk flow. Self-gravity \citep{toomre64,kratter16}, the magnetorotational instability \citep{balbus91} as (strongly) modified by non-ideal magnetohydrodynamic effects, and the Vertical Shear Instability \citep{nelson13} are the leading suspects. These processes generate turbulence, which has distinctive (and potentially observable) characteristics on large scales.
    \item Loss of mass and / or angular momentum through disk winds. Photoevaporation \citep{hollenbach94} -- in which the surface of the disk is heated by high-energy photons to the point where it escapes in a thermal wind -- is guaranteed to occur at some level \citep{alexander14,ercolano17}. It is a pure mass-loss process. Magnetized winds \citep{blandford82,pudritz83,bai13}, on the other hand, remove gas that has {\em more} angular momentum than the disk at the launch point, simultaneously dispersing the disk while driving accretion \citep{suzuki09,armitage13}
\end{itemize}
Figure~\ref{Armitage_evolution_figure} illustrates these possibilities. In addition to global evolution, many of these processes can also plausibly generate rings, spirals, and other sub-structures in disks.

The state-of-the-art in direct numerical simulation \citep{xu21,mauxion24} allows disks to be followed for about $10^5 \ {\rm yr}$, roughly the duration of the Class~0 phase of Young Stellar Object evolution. Longer duration evolution, and studies that seek to compare theory against observed populations of disks, necessarily require simpler models. The minimal goal of one-dimensional models is to predict $\Sigma(r,t)$, though it is also possible to evolve the disk central temperature \citep[this is essential for modeling disk outbursts;][]{bell94,armitage01,zhu09}. The vertical magnetic flux threading the disk, $B_z(r,t)$, is another important dynamical quantity that in principle should be evolved in models that include magnetized winds or the magnetorotational instability. The simplest case is a disk that evolves solely due to local turbulent redistribution of angular momentum. The appropriate one-dimensional description is a diffusion equation \citep{lyndenbell74,pringle81},
\begin{equation}
    \frac{\partial \Sigma}{\partial t} = \frac{3}{r} 
     \frac{\partial}{\partial r} \left[
     r^{1/2} \frac{\partial}{\partial r} \left( \nu \Sigma r^{1/2} \right) \right],
\label{eq_diffusion}     
\end{equation}
in which $\nu$ is an effective viscosity that is generated by turbulence. If $\nu$ is approximated by a time-independent power-law function of radius, this equation admits a similarity solution (taking the form of a time-varying exponentially truncated power-law for the surface density) that has been widely used for disk modeling \citep[e.g.][]{hartmann98}. More generally, the viscosity can be written in terms of the local scale height $h$ and disk sound speed $c_s$ as \citep[][the ``$\alpha$-prescription"]{shakura73},
\begin{equation}
    \nu = \alpha c_s h,
\end{equation}
with $\alpha$ being a dimensionless parameter that characterizes the efficiency of angular momentum transport.

Wind-driven evolution can be modeled within an analogous framework, by adding sink and advective terms to the right hand side of equation~(\ref{eq_diffusion}) that represent the effects of mass and angular momentum loss on the underlying disk. The detailed form of the additional terms depends on the choice of wind model. \cite{tabone22}, by introducing an ``$\alpha$-like" parameter for the wind torque, derived an analytic generalization of the \cite{lyndenbell74} solution for one choice that captures the generic effects of combined viscous / wind-driven disk evolution. \cite{chambers19} derived a solution that includes the effects of irradiation and viscous heating on the evolving disk structure. These analytic models are useful for rapid population modeling. The effects of infall of gas onto the disk \citep{lin90}, and the long timescale evolution of the net magnetic flux \citep{lubow94}, also have one-dimensional descriptions, though the more complex physics of these processes is harder to capture reliably in a reduced model.

Determining the relative importance of these evolution processes, from either theoretical or observational arguments, is a major challenge. It has often been assumed that the ordering from top to bottom in Figure~\ref{Armitage_evolution_figure} maps onto a {\em temporal sequence}, in which disks form from infall on a time scale of $\sim 10^5 \ {\rm yr}$, evolve initially as massive self-gravitating structures and then more slowly under the action of viscosity and winds, and are finally dispersed by photoevaporation. This remains a plausible paradigm. Observations of molecular line broadening \citep{flaherty17}, and of the thickness of dust layers \citep{villenave22,villenave25}, imply low levels of turbulence in many \citep[though not all;][]{flaherty24} of the best-studied disks. The inferred values of $\alpha$ in the ``weakly turbulent" disks -- typically $\alpha \sim 10^{-4}$ or lower -- are broadly consistent with strongly damped MHD turbulence \citep[described as ``dead zone" models;][]{gammie96} and weak purely hydrodynamic driving mechanisms. As a consequence, magnetized disk winds rather than turbulence are considered to be dominant in most recent models. There remains room, however, for larger deviations from conventional wisdom, in particular models in which infall is dynamically important at late as well as early times. Indeed, simulations of star formation \citep[reviewed by][]{kuffmeier24} have long found that encounters and infall drive a highly dynamic mode of disk evolution \citep{seifried13,bate18}.

A primary result of ALMA (Atacama Large Millimeter/submillimeter Array) observations has been the discovery that protoplanetary disks typically show pronounced {\em substructure}, in the form of departures from the simplest model of an axisymmetric disk with a monotonically decreasing profile of intensity \citep{andrews18}. Substructure can be detected most economically in dust continuum imaging, but is also seen in molecular line data that directly traces the gas. Rings \citep{HL_Tau_alma15}, crescent-shaped features \citep{vandermarel13}, and spirals \citep{perez16} are all observed, roughly in decreasing order of frequency \citep[for reviews, see][]{andrews20,bae23}. Substructure is also detected in scattered light images of disks \citep{avenhaus18}, most prominently around more massive stars. Scattered light substructure arises from small grains in the disk atmosphere, and is often harder to interpret than sub-mm images that trace regions closer to the mid-plane. Resolution limitations mean that all of the {\em observed} substructure involves orbital radii and spatial scales $r > {\rm au}$; typically the impact on planet formation would be in the outer disk at $r > 10$~au, well beyond the water snowline. Different physical processes may very well lead to the existence of cavities and rings at much smaller, sub-au scales, including from the interaction of a stellar magnetosphere with the disk \citep{zhu24} and from the onset of stronger turbulence due to the magnetorotational instability at temperatures where the alkali metals are thermally ionized \citep{gammie96}. These structures could be central to our understanding of close-in planet formation \citep{chatterjee14}.

The proximate cause of currently observed substructure in dust continuum images is understood to be the aerodynamic concentration of particles in local gas pressure maxima (or, in weaker cases, local pressure enhancements over a smooth background). This process -- often described imprecisely as ``trapping" -- was foreseen by \cite{whipple72} (look at his amazingly prescient Figure~1). The pressure maxima can be in form of axisymmetric rings \citep{haghighipour03}, as in the original Whipple proposal, vortices \citep{barge95}, or spiral waves \citep{rice04}, matching the main classes of observed structures. Ring-like dust structures are also predicted to form without the need for underlying gas perturbations in the vicinity of condensation / sublimation fronts \citep{zhang15}, but this effect does not appear to be responsible for most of the substructure that is currently being seen.

While the immediate cause of dust substructure is reasonably clear, the same cannot be said for why gas disks develop pressure maxima in the first place. A popular hypothesis is that the source of substructure is gravitational perturbations from already-formed planets. Planets launch spiral density waves, and open gaps  whose edges -- which are approximately axisymmetric pressure maxima -- can become unstable to vortex generation through the Rossby Wave Instability \citep{lovelace99}. Planet-disk interaction models generate disk substructure that is often a good match to what is actually observed  \citep{zhang18}, and are surely responsible for {\em some} substructure. That said, simulations of disk turbulence and magnetized winds also commonly lead to the formation of rings \citep{johansen09,simon14,suriano18,riols19} and vortices, while self-gravity of course forms spirals. Infall can also generate substructure \citep{kuznetsova22}. As a result, the interpretation of substructure is uncertain. For the purposes of this review, the key takeaway is that the existence of substructure has profound implications for planet formation {\em irrespective of the interpretation}. If ringed disks almost always host planets, that means that a substantial population of ice or gas giant planets is able to form surprisingly quickly at orbital separations that are as large as $\sim 50~{\rm au}$. If substructure is instead a byproduct of disk transport processes, then the observed concentration of solids that results modifies the evolution of all subsequent phases of planetary growth. 

\section{Aerodynamic and collisional evolution of solids}
\begin{figure}[t]
\begin{center}
\includegraphics[width=0.85\textwidth]{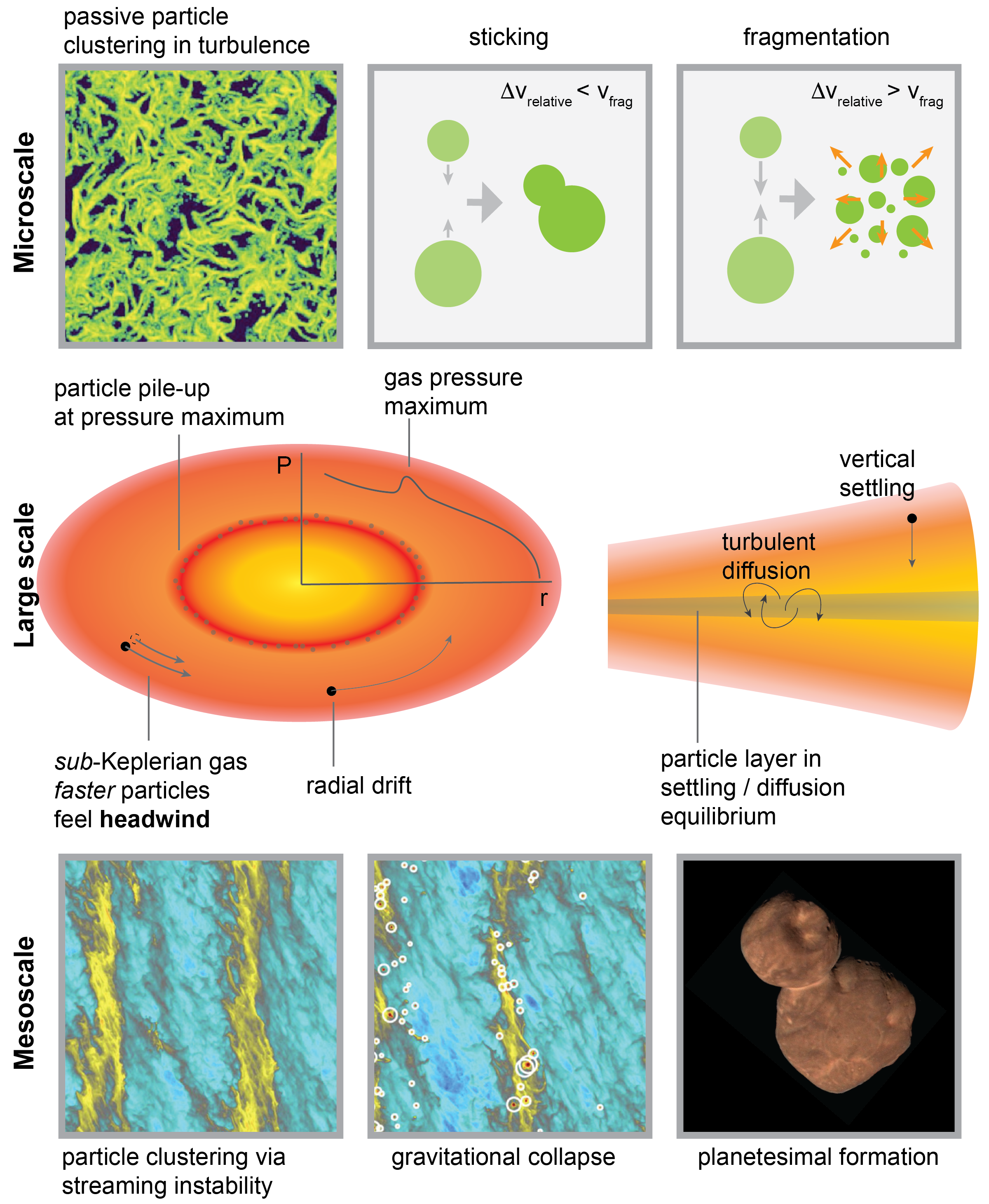}
\end{center}
\caption{Aerodynamic coupling between solid particles and disk gas controls the early phases of planet formation on multiple scales. On the smallest scale, it leads to clustering of solids \citep[simulation from][]{chan24}. The characteristic clustering scale is the Kolmogorov scale defined for isotropic fluid turbulence -- of the order of km at 5~au \citep{pan11}. It also results in collisions between individual particles \citep{ormel07}. On intermediate scales, it leads to the streaming instability \citep{youdin05} and planetesimal formation \citep[simulation in a $(0.2h)^3$ domain from][]{simon16}. The Kuiper Belt contact binary Arrokoth (bottom right panel) is consistent with this formation channel. On large, observable scales, aerodynamics leads to vertical settling, radial drift, and particle concentration at pressure maxima.}
\label{Armitage_aerodynamic_figure}
\end{figure}

Dust inherited by the disk is expected to follow the size distribution, $n(a) \propto a^{-3.5}$, that is inferred from extinction measurements in the interstellar medium \citep{mathis77}. Growth up to the scale of small macroscopic solids (``pebbles", which might be anywhere between mm and meters in physical size) is controlled by aerodynamic and collisional processes. The physics of these processes, reviewed by \citet{birnstiel24}, is quite subtle, but most of the important results follow from two considerations: 
\begin{enumerate}
    \item The measured material properties of silicate and water ice aggregates imply that fragmentation occurs when collision velocities exceed a threshold value (nominally 1~ms$^{-1}$, though possibly up to 10~ms$^{-1}$ for ice in some circumstances). Collision speeds in a turbulent gas increase with particle size, so this material limit translates to a predicted {\bf maximum physical particle size}, normally in the mm-cm range.
    \item Solid particles are coupled to the gas via aerodynamic drag forces, but imperfectly so. Deviations from perfect coupling have multiple consequences, including vertical settling, radial drift, particle concentration in vortices, and small-scale clustering and collisions. Back-reaction of the aerodynamic forces on the gas, when the dust to gas density ratio is non-negligible, leads to additional effects, including most notably the streaming instability. These effects depend not on the physical particle size, but on a dimensionless measure of ``size" for aerodynamic purposes called the {\bf Stokes number} (equation~\ref{eq_Stokes_number}). 
\end{enumerate}
Figure~\ref{Armitage_aerodynamic_figure} illustrates some of these processes, which operate all the way from observable global scales down to microscopic scales on which dust particles are colliding.

Small dust grains in gaseous disks experience low velocity collisions due to Brownian motion \citep{dullemond05}, that lead to sticking and the formation of porous aggregates. Larger particles collide at higher velocities, due to coupling to small-scale turbulence or size-dependent drift, leading to compactification and, eventually, bouncing or fragmentation. A microscopic theory \citep[``JKRS theory", applied to the astrophysical context by][]{dominik97} relates the collision outcome to the surface energies of the materials involved, and this theory can be used to model the growth of aggregates via molecular dynamics techniques \citep[e.g.][]{wada09}. The fiducial results for grain growth, however, come from laboratory experiments, usually performed in microgravity conditions \citep{blum08}. For silicate particles, considered to be representative of the inner disk interior to the snowline, net growth occurs for collisions with relative velocity $\Delta v < 1 \ {\rm ms}^{-1}$ \citep{blum08}. For water ice, collision experiments by \cite{gundlach15} at $T \approx 100-200 \ {\rm K}$ found growth for $\Delta v < 10 \ {\rm ms}^{-1}$.  Measurements of water ice surface energies, however, are similar to values measured for silicates at temperatures $T < 175 \ {\rm K}$ \citep{musiolik19}. In JKRS theory this would imply that both silicate and ice aggregates cease to stick above similar (low) threshold collision speeds, except near the snowline where water ice unambiguously shows a higher threshold. There is evidently some uncertainty in the sticking efficiency of water ice aggregates, and it is common in models to consider the sticking and fragmentation thresholds for ice as parameters and explore the effect of different values in the 1-10~ms$^{-1}$ range.

Translating the experimental results for the sticking and fragmentation thresholds into a predicted maximum particle size $a_{\rm max}$ requires knowledge of what sets $\Delta v$ as a function of size $a$. In the most common situation, collisions at $a \sim a_{\rm max}$ are caused by aerodynamic coupling to small-scale turbulence \citep[this process also leads to small-scale clustering of solids;][]{cuzzi01}. For particles of size $a_1$ and $a_2$, coupled aerodynamically to isotropic Kolmogorov turbulence, \cite{ormel07} provide the standard (astrophysical) calculation that gives $\Delta v (a_1,a_2)$. Adopting an $\alpha$ model for the turbulence, the peak collision speed is found to be,
\begin{equation}
    (\Delta v)_{\rm max} \sim \alpha^{1/2} c_s,
\end{equation}
where $c_s$ is the sound speed. Noting that $c_s = h \Omega_{\rm K}$, and taking $\alpha = 10^{-4}$ and $h/r =0.1$, this result implies that $(\Delta v)_{\rm max} > 1 \ {\rm ms}^{-1}$ wherever the orbital velocity exceeds 1~kms$^{-1}$ (for a Solar mass star, out to almost 1000~au). One concludes that the material properties of particles often impose a strong constraint on growth. Detailed calculations (including radial drift, which is usually a stronger limiter at large radii) show that the maximum size of particles that form from pairwise collisional evolution is perhaps 0.1~mm at 100~au, increasing to around 1~cm at 3~au \citep{birnstiel24}. 

Aerodynamic processes in disks depend on the {\em stopping time} of particles. A particle of mass $m$, moving at speed $\Delta v$ relative to the gas, feels an aerodynamic drag force $F_{\rm drag}$ and slows down on a time scale,
\begin{equation}
    t_{\rm stop} = \frac{m \Delta v}{| F_{\rm drag} |}.
\label{eq_tstop}    
\end{equation}
It turns out to be very useful to multiply $t_{\rm stop}$ by a characteristic frequency and make the stopping time dimensionless. In isotropic turbulence (relevant to the grain-grain collisions discussed above) the appropriate characteristic frequency is that of eddies, but for most other applications the orbital angular frequency is the right choice. This is called the Stokes number (or, the dimensionless stopping or friction time),
\begin{equation}
    \tau_{\rm S} \equiv t_{\rm stop} \Omega.
\label{eq_Stokes_number}    
\end{equation}
The form of the drag force depends on particle size \citep{weidenschilling77}. For $a < \lambda$, the mean free path of molecules in the gas, we are in the Epstein regime, and for a spherical particle,
\begin{equation}
 {\bf F}_{\rm drag} = - \frac{4 \pi}{3} \rho a^2 v_{\rm th} \Delta {\bf v}.
\end{equation}
Here, $\rho$ is the gas density and $v_{\rm th}$ its thermal speed (roughly equal to the sound speed). With this drag law, the stopping time is independent of $\Delta v$ and the dimensionless Stokes number is fixed by the size and composition of the particle. In the inner disk, the gas density is high enough that particles near the fragmentation limit have $a > \lambda$. This is the Stokes regime, in which the gas flows as a fluid around the particle and $F_{\rm drag} \propto (\Delta v)^2$. \cite{cheng09} provides a parameterization of the drag force that is valid across both regimes.

In the vertical direction, particles settle toward the disk midplane at approximately their terminal velocity. Settling is opposed by turbulent diffusion. In the tracer limit (i.e. assuming that the volume density of dust is negligible compared to the gas), and assuming that the magnitude of turbulent diffusion (in the vertical direction) equals the viscosity, $D=\nu$, the equilibrium particle distribution is a gaussian \citep{dubrulle95}. The scale height of the particles, $h_{\rm d}$, is related to that of the gas through \citep{birnstiel24},
\begin{equation}
    \frac{h_{\rm d}}{h} = \sqrt{\frac{\alpha}{\tau_{\rm S}+\alpha}}.
\end{equation}
This allows observations of the thickness of dust disks to be used to constrain disk turbulence \citep{villenave25}. The assumption that $D=\nu$ is quite crude, and there can be substantial deviations from this simple formula in disks where the turbulence is driven by the Vertical Shear Instability \citep{stoll16,pfeil24,fukuhara24} or by self-gravity \citep{baehr21}.

In the radial direction, aerodynamic effects lead to drift of particles relative to the gas. For $\tau_{\rm S} \ll 1$, the strong aerodynamic coupling means that dust orbits at almost the same speed as the gas, but, unlike the gas, it {\em does not feel the pressure gradient}. The unbalanced radial force leads to drift in the direction of higher gas pressure, which in a smooth disk would be inward. The general expression for the drift speed, valid not just in the trace particle limit, was derived by \cite{nakagawa86}. If the gas orbits at speed, 
$v_\phi = (1 - \eta) v_{\rm K}$, then dust with a dust to gas ratio $\epsilon = \rho_{\rm dust} /\rho$ drifts radially at speed \citep[e.g.][]{birnstiel24},
\begin{equation}
    v_{\rm r,dust} = - \frac{2 \eta v_{\rm K}}{\tau_{\rm S}+(1+\epsilon)^2 \tau_{\rm S}^{-1}}.
\end{equation}
Even though $\eta \sim (h/r)^2$ is small, typically only $\sim {\rm few} \times 10^{-3}$, the rate of the resulting drift is highly consequential. At 1~au, the time scale for drift of particles with $\tau_{\rm S} = 1$ is as short as of the order of a thousand years. Particles with this Stokes number would be of the order of meters in size, leading to the moniker ``meter-sized barrier" for this aerodynamic constraint on planetesimal formation. (Material physics limits on sticking already make the growth of meter-sized boulders improbable, but the meter-sized barrier provides an independent, and arguably more robust, argument against planetesimal formation by pairwise coagulation.) More importantly in a modern context, particles with physical sizes of mm-cm and Stokes numbers $\tau_{\rm S} \sim 10^{-2}$, that can readily form given our best understanding of coagulation physics, are predicted to suffer substantial drift even in the outer disk on scales of 10-100~au. 

In disks with dynamically insignificant substructure, radial drift is inward. It leads to an effective limit on particle size: at any radius, particles cannot grow to a size for which their growth time scale exceeds the radial drift time scale \citep{birnstiel12}. Defining $\gamma = | \partial \ln P / \partial \ln r |$, the maximum particle size in drift-limited growth is \citep{birnstiel24},
\begin{equation}
    a_{\rm max,drift} = \frac{2 \Sigma_{\rm dust}}{\pi \rho_{\rm s} \gamma} \left( \frac{h}{r} \right)^{-2},
\end{equation}
where $\Sigma_{\rm dust} (r)$ is the dust surface density and $\rho_{\rm s}$ is the material density. This limit can be compared to the limit imposed by fragmentation \citep{birnstiel24},
\begin{equation}
    a_{\rm max,frag} = \frac{2 \Sigma}{3 \pi \rho_{\rm s} \alpha} \left( \frac{\Delta v_{\rm frag}}{c_s} \right)^2.
\end{equation}
The drift limit is usually the most stringent in the outer disk, while fragmentation limits growth closer to the star.

Turbulent diffusion in the radial direction can be important both for the evolution of trace gas species, and for the evolution of dust populations that are simultaneously subject to radial drift \citep{morfill84,clarke88}. \cite{tominaga19} derive formulae for dust diffusion in disks that conserve angular momentum, and which can be used to model dust concentration processes and possible instabilities. An important consequence of diffusion is the qualitative change it makes to how dust evolves in the vicinity of gas pressure maxima. Ignoring diffusion, an axisymmetric pressure maximum -- whether it forms spontaneously or at a planetary gas edge -- is efficient at trapping aerodynamically drifting solids, filtering all but the smallest particles that are swept inward by the mean flow \citep{rice06}. With diffusion, it is always possible to find a  steady state solution in which the turbulent flux overcomes the aerodynamic trapping effect, such that pressure maxima temporarily detain but do not permanently imprison solids \citep{ward09}. Numerical simulations confirm that so-called ``particle traps", are, generically, rather leaky \citep{zhu12,huang25}.

\cite{birnstiel24} reviews dust growth in protoplanetary disk in much more detail than can be covered here. DustPy \citep{stammler22} is open source software that can be used to calculate how solids grow and drift within disks.

\section{Planetesimal formation}

{\em Planetesimals} are primordial bodies whose sizes, although not precisely defined, are usually considered to be in the range between 100~m and a few hundred km. The growth of planetesimal-scale bodies is a key milestone in planet formation because it marks the double onset of gravity's dominance. Above some critical size, probably around a few hundred meters \citep{benz99}, self-gravity rather than material strength controls the outcome of physical collisions. Above some other, rather larger physical size, gravitational forces become more important than aerodynamic ones in the dynamics of a population of bodies orbiting within the disk. 

Theories on how planetesimals form have shifted markedly over time. \cite{goldreich73}, and independently \cite{safronov72}, suggested that dust settling proceeds until the mid-plane dust density exceeds the Roche density, which at orbital radius $r$ is,
\begin{equation}
    \rho_{\rm Roche} \sim \frac{M_*}{r^3}.
\end{equation}
If $\rho_{\rm dust} > \rho_{\rm Roche}$, solid material can collapse gravitationally to form planetesimals. Although there are circumstances where this process may operate, it is no longer considered to be a generally viable planetesimal formation mechanism. Typically, vertical shear will generate enough turbulence to halt settling before the critical density is reached, even in an {intrinsically non-turbulent} disk \citep{cuzzi93}. The other obvious mechanism for planetesimal formation is a continuation of pairwise sticking collisions from pebble to planetesimal scales \citep{weidenschilling93}. While this model cannot be entirely ruled out, it is largely unviable given current understanding of fragmentation and radial drift processes. 

The current working hypothesis for planetesimal formation is that the {\em streaming instability} \citep{youdin05} generates particle clumps that are dense enough to collapse gravitationally to planetesimals. The streaming instability is a local, linear instability of the \cite{nakagawa86} solution for equilibrium radial drift. It requires consideration of the momentum feedback particles exert on the gas, and a non-zero drift velocity (hence, it cannot formally be operative at the peak of a pressure maximum). \cite{youdin05} show that the streaming instability has non-zero (though sometimes small) growth rates for almost all combinations of Stokes number and dust to gas ratio, and it should therefore be generically present in protoplanetary disks. \cite{jacquet11} and \cite{magnan24} discuss the rather subtle physical mechanism of the streaming instability, which is the most important (for planet formation) of a broader class of resonant drag instabilities \citep{squire18}. It should be noted that the \cite{nakagawa86} equilibrium studied by \cite{youdin05} differs in multiple ways from both real disks and from the setups for most streaming instability simulations. The \cite{nakagawa86} model is vertically unstratified (whereas, in actual disks, vertical gravity is not normally negligible), does not include turbulent diffusion (which is present if there is any instrinsic disk turbulence and which is needed to establish a stratified particle layer equilibrium), and is based on a fluid description of the particles (this is only a good approximation for $\tau_{\rm S} \ll 1$, and is not used in particle-based simulations). Subsequent linear calculations have included the effects of a particle size distribution \citep{krapp19,paardekooper20,zhu21} and turbulent diffusion \citep{chen20,umurhan20,gerbig24}. Both effects act to reduce the growth rates and restrict the range of parameters over which linear instability is present.

Numerical simulations, starting with the pioneering work of \cite{youdin07} and \cite{johansen07}, have demonstrated a streaming-initiated pathway to planetesimal formation. It is illustrated in Figure~\ref{Armitage_aerodynamic_figure} \citep[using simulations from][]{simon16}. From uniform initial conditions, the streaming instability generates largely axisymmetric filaments of particles on small scales of the order of $\sim 0.1 h$. Such structures may have observable effects on disk emission \citep{scardoni21}, but they are not presently able to be directly imaged. (They are {\em not} the rings seen in ALMA disk images.) Within the filaments dense clumps form, and {\em given favorable conditions} these clumps can exceed the Roche density and collapse gravitationally. The expected outcome of collapse, which cannot be followed to material density with current simulation techniques, is some combination of single planetesimals, binaries \citep{nesvorny10,nesvorny21}, or small clusters of planetesimals.

\cite{simon22} review planetesimal formation physics, and the large simulation literature. One important set of results concerns the threshold dust to gas ratio needed for planetesimal formation, defined via a metallicity $Z = \Sigma_{\rm dust} / \Sigma$. \cite{carrera15} found that the critical metallicity for ``prompt" (i.e. basically dynamical time scale) planetesimal formation, as a function of the Stokes number, follows a U-shaped curve with a minimum near $\tau_{\rm S} \simeq 0.1$. Subsequent revisions to the critical curve \citep{yang17,li21,lim24} show that for a standard value of the pressure gradient, $Z_{\rm crit}$ is below 0.01 -- the fiducial value of the global dust to gas ratio in disks -- for $0.01 \leq \tau_{\rm S} \leq 1$. These thresholds apply to the most favorable scenario of negligible intrinsic disk turbulence and a monodisperse particle size distribution. When turbulence is present, the criterion for planetesimal formation is approximately that the settled particle layer has a mid-plane density $\rho_{\rm dust} / \rho > 1$ \citep{gole20,lim24a}.

A second important simulation result is the predicted mass function of planetesimals formed from the streaming instability. A large number of studies agree that streaming-initiated collapse leads to a top-heavy initial planetesimal mass function, in which most of the total mass is in the largest objects \citep{johansen07,simon16,schafer17,li19,schafer24}. The absolute size of the largest planetesimals is of course a function of the disk conditions, but for reasonable parameters it is often hundreds of kilometers. This first principles prediction is {\em large} compared to what was assumed in most formation-agnostic planetary growth models in the past, and has substantial downstream implications both for the Solar System \citep[``asteroids were born big";][]{morbidelli09} and for planet formation generally. In particular, planetesimals that are 100~km and larger in size experience much less aerodynamic drag than 0.1-1~km planetesimals, and when their orbits are excited by gravitational perturbations the population will saturate at a higher velocity dispersion. This {\em reduces} the rate at which protoplanets grow by planetesimal accretion, so that, counter-intuitively, being able to rapidly make {\em large} primordial planetesimals does not necessarily imply faster overall planet formation time scales.

There is no analytic theory for the form of the initial planetesimal mass function, but several simple functions (an exponentially tapered power-law, a broken power-law, etc) provide fits to numerical results that can be compared to Solar System populations of small bodies. The Kuiper Belt is widely considered to be the most primordial surviving population and offers the greatest potential for a test, though even there careful consideration has to be given to the extent of collisional evolution \citep{morbidelli21,benavidez22}. Other possible tests of the predictions of planetesimal formation models include the properties and inclination distribution of Kuiper Belt binaries \citep{grundy19,porter24}, which are consistent with streaming instability expectations \citep{nesvorny19}. The detailed morphology of the single small Kuiper Belt Object imaged by the {\em New Horizons} mission \citep[Arrokoth;][]{mckinnon20} also supports a streaming formation mechanism.

Given the limitations of both the Solar System data and of current numerical simulations, tests of the streaming hypothesis for planetesimal formation are not decisive. Other proposals for planetesimal formation include collisional growth of porous icy aggregates \citep{okuzumi12}, and various non-streaming processes that can concentrate solids to above the threshold for gravitational collapse while avoiding the barriers that hinder the Goldreich-Ward mechanism. Examples include secular gravitational instability \citep{ward76,youdin11,takahashi14}, and particle concentration in spiral arms \citep{rice04,baehr22,longarini23} or vortices \citep{lyra24}. Secular gravitational instability is interesting, in part, because the favorable range in Stokes number may be smaller than for streaming instability, while models that invoke gravitational collapse of particle clumps in disk substructures tend to produce extremely massive ``planetesimals" that would more commonly be considered to be planetary embryos or dwarf planets.

\section{Formation of rocky planets, cores, and giants}
\label{Armitage_sec_planet_growth}

\begin{figure}[t]
\centering
\includegraphics[width=0.85\textwidth]{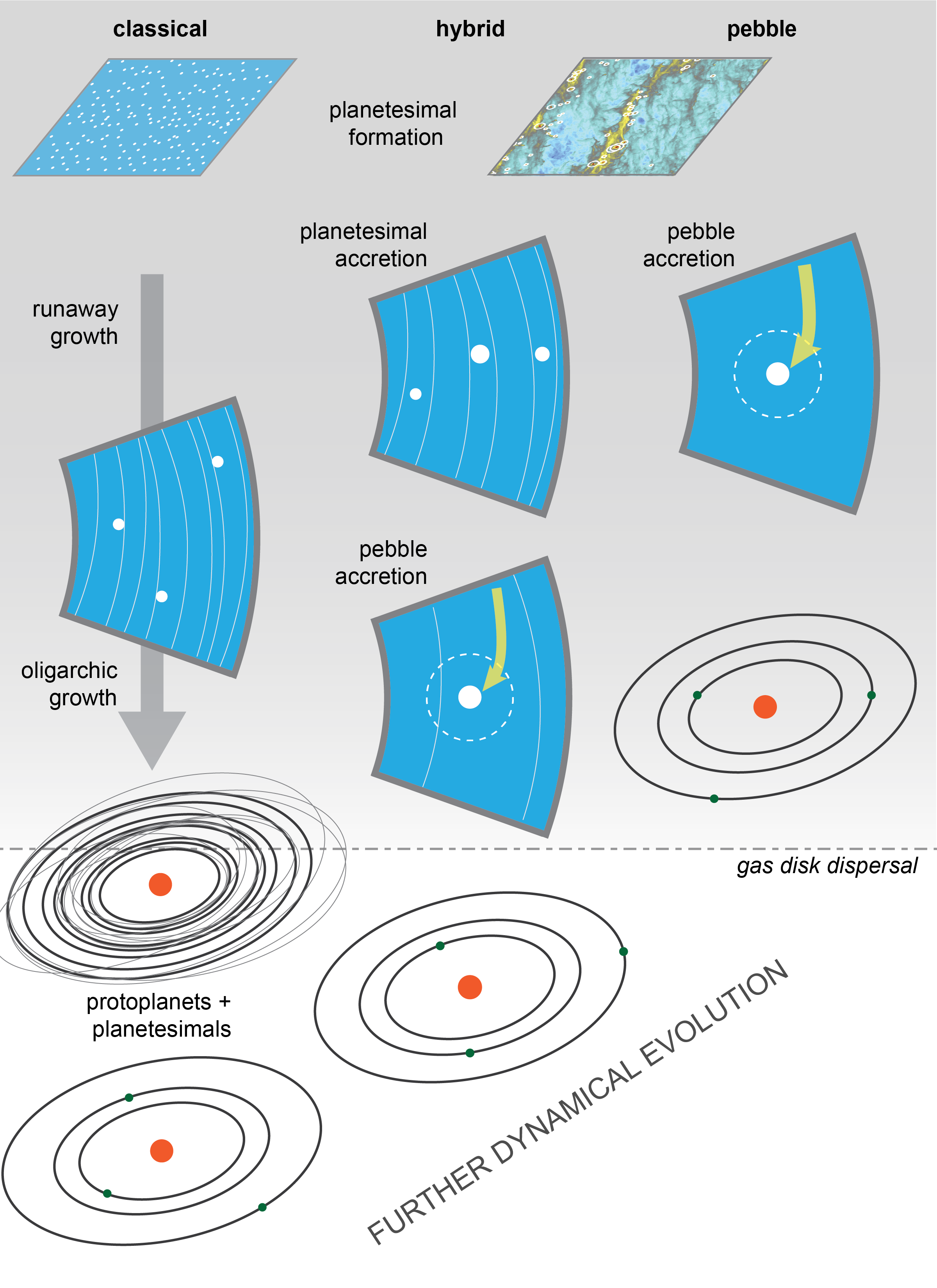}
\caption{Rocky planets (including the Solar System's terrestrial planets and low mass exoplanets), and the cores of more massive planets, grow from planetesimals by a combination of planetesimal and pebble accretion. The classical model for terrestrial planet formation invokes only planetesimal growth, while more recent ``hybrid" models involve an initial phase of planetesimal growth that is followed by predominantly pebble accretion. If the most massive planetesimals are large enough, a pure pebble model for subsequent growth may also be viable.}
\label{Armitage_fig_pebbles_planetesimals}
\end{figure}

Once planetesimals have formed, their initial growth to become protoplanets, and ultimately either rocky planets or the cores of ice or gas giants, occurs via two physically distinct mechanisms:
\begin{enumerate}
    \item {\em Planetesimal accretion}. Planetesimals physically collide, due to gravitational interactions, at low enough velocity to grow. The rate of collisions is determined, through the physics of gravitational focusing, by the velocity dispersion of the planetesimals. To a first approximation, one imagines the protoplanet growing from a local planetesimal reservoir that does not drift radially.
    \item {\em Pebble accretion}. While gas is still present, a fraction of the appropriately aerodynamically coupled pebbles (typically, those with $10^{-3} \leq \tau_{\rm S} \leq 1$) that enter the protoplanet's Hill sphere are gravitationally captured and accreted via a settling flow. The protoplanet accretes a fraction of the solids that drift past its orbit, with an efficiency that depends on its mass, on the Stokes number, and on the disk properties.
\end{enumerate}
Figure~\ref{Armitage_fig_pebbles_planetesimals} illustrates the three ways these mechanisms could combine for an overall scenario of planetary growth. The classical (pre-2010) model considers only planetesimal growth. This must be wrong at some level -- as we observe roughly pebble-sized solids in disks -- but planetesimal-driven growth could still be dominant at some times or locations in the disk. The other extreme is pure pebble growth. This is conceivable, but requires that at least some very massive planetesimals (probably, more massive than one would expect from the streaming instability) form. The most popular model is a hybrid, in which the earliest phases of growth occur via planetesimal collisions, with pebble accretion contributing and in some regimes dominating once larger protoplanets have formed. In any of these scenarios: planetesimal, hybrid, or pebble, further evolution occurs due to orbital instabilities (planet-planet collisions) and due to the accretion of envelopes by sufficiently massive cores.

The key concepts of planetesimal-driven growth can be illustrated with a toy 
model for planet growth within an initially dynamically cold planetesimal disk. Consider a protoplanet of 
mass $M$, radius $R$, and surface escape speed $v_{\rm esc}$, orbiting within a 
disk of planetesimals with surface density $\Sigma_p$ and velocity dispersion 
$\sigma$. The typical eccentricity and inclination of the planetesimals are  
$e \sim i \sim \sigma / v_K$, so the planetesimal disk has thickness $\sim \sigma / \Omega$. 
Simple collision rate arguments \citep[e.g][]{lissauer93,armitage20} give a growth rate,
\begin{equation}
 \frac{{\rm d}M}{{\rm d}t} = \frac{\sqrt{3}}{2} \Sigma_p \Omega \pi R^2 
 \left( 1 + \frac{v_{\rm esc}^2}{\sigma^2} \right),
\label{eq_pib} 
\end{equation}
that describes growth in the so-called ``dispersion dominated" limit (there is also a 
shear-dominated limit when tidal effects are dominant). The enhancement of the physical 
collision cross-section due to the effect of {\em gravitational focusing} is captured by 
the term in parenthesis. It can 
vary by several orders of magnitude, such that growth in the classical picture is closely 
tied to the evolution of $\sigma$. There are two regimes,
\begin{itemize}
\item
{\em Runaway growth} \citep{greenberg78}. A small protoplanet does not affect the local planetesimal velocity dispersion 
(typically, $\sigma$ is determined by balancing excitation from  
planetesimal-planetesimal scattering encounters against aerodynamic damping). 
Assume that $\sigma = {\rm const} \ll v_{\rm esc}$. Then for two bodies in the 
same disk environment, with masses $M_1 > M_2$, equation~(\ref{eq_pib}) predicts that  
${\rm d} (M_1/M_2) / {\rm d}t \propto (M_1/M_2)(M_1 ^{1/3} - M_2^{1/3} ) > 0$. Growth 
amplifies any initially small mass differences. 
\item
{\em Oligarchic growth} \citep{kokubo98}. More massive protoplanets excite the velocity dispersion of 
nearby planetesimals, slowing their own growth while less massive neighbors catch up. This leads to a population of {\em planetary embryos} growing at comparable rates.
\end{itemize}
Ignoring migration the outcome is a system of 
protoplanets on low-eccentricity orbits, which begin to collide among themselves when their mass becomes comparable to the mass in the sea of remaining planetesimals \citep{kenyon06}. This end-point of planetesimal growth has traditionally been used as the effective initial condition for N-body simulations of the final assembly phase of rocky planets. Although these planetesimal-only models are rather simple -- arguably too simple -- they are quite successful in reproducing the basic properties of the Solar System's terrestrial planets \citep{raymond09}.

Pebble accretion \citep{ormel10,lambrechts12} is a qualitatively different physical process for planetary growth. While gas is present, small solid particles that encounter the planet within its Hill sphere are accreted if the time scale on which they would settle toward the planet is shorter than the time scale of the encounter. The accretion rate is particularly simple to estimate in the limit where the thickness of the pebble layer is larger than the Hill sphere of the planet,
\begin{equation}
    r_{\rm Hill} = \left( \frac{M}{3M_*} \right)^{1/3} r.
\end{equation}
Within the Hill sphere, by definition, the planet's gravity dominates over the tidal gravitational field of the star, and it makes sense to define a time scale on which a pebble would settle toward the planet. Equating the drag force (equation~\ref{eq_tstop}) to the gravitational force for a pebble encountering the planet at impact parameter $b$ and relative velocity $\delta v$, the terminal velocity with which settling occurs is,
\begin{equation}
    v_{\rm settle} = \frac{GM t_{\rm stop}}{b^2}.
\end{equation}
Applying the condition that the settling time scale $t_{\rm settle} = b / v_{\rm settle}$ equals the encounter time scale, $t_{\rm enc} = b / \delta v$, we find that $\delta v = GM t_{\rm stop} / b^2$, and the pebble accretion rate is,
\begin{equation}
 \dot{M} \sim \rho_{\rm pebble} b^2 \delta v \sim GM \rho_{\rm pebble} t_{\rm stop},
\end{equation}
where $\rho_{\rm pebble}$ is the density of the pebble ``fluid". This is an approximation for the pebble accretion rate in what is described as the three dimensional regime. Similar estimates can be derived in the two dimensional limit, where the pebble layer is thinner than the Hill sphere. In this regime, the accretion rate varies depending on whether the encounter velocity is due to disk shear, the headwind experienced by the pebbles as they drift, or a non-zero planetary eccentricity. Detailed expressions are given in \citet{liu18} and \citet{ormel18}. Generalizations to the case of a polydisperse size distribution of solids are given in \citet{lyra23}.

Pebble accretion has several distinctive differences from planetesimal accretion. There is no regime of pebble accretion that exhibits runaway growth. It is also fairly strictly bounded in planet mass. Pebble accretion is inefficient below a {\em pebble initiation mass} \citep{visser16}, and for typical parameters would only be able to directly grow planets from planetesimals if their initial sizes are rather large (at least several hundred km). Pebble accretion is also likely to be inefficient above a {\em pebble isolation mass} \citep{lambrechts14},
estimated by \citet{bitsch18} as,
\begin{equation}
    M_{\rm iso, pebble} \simeq 25 \left( \frac{h/r}{0.05} \right)^3 
    \left( \frac{M_*}{M_\odot} \right) f M_\oplus,
\end{equation}
where $f$ is a correction factor that depends on the radial pressure gradient, viscosity, and strength of turbulent diffusion (see Chapter by Ormel in this volume). Note the strong dependence on $(h/r)$. Above this mass, a growing planet starts to open a gap in the disk, forming a local pressure maximum that can stop the radial drift of solids. Simulations show that this suppression of radial drift can be reduced by three dimensional effects \citep{huang25}.

Planetesimal accretion is severely limited at large orbital radii by the tendency of more massive bodies to scatter, rather than accrete, planetesimals \citep[this is the origin of the long-known difficulty in forming Uranus and Neptune in situ from planetesimals;][]{levison01}. Pebble accretion, conversely, tends to work {\em better} at moderately large orbital radii -- solids at the drift or fragmentation limit have more favorable aerodynamic properties, and there is no reduction in accretion efficiency as the ratio of Hill sphere size to planetary radius increases. Outside perhaps 5-10~au, pebble accretion is presumptively the dominant mechanism for planetary growth \citep[e.g.][]{drazkowska23}. More general statements as to the relative role of planetesimals and pebbles in planet formation are, however, hazardous. Planetesimal and pebble accretion are distinct processes that cannot be easily compared, and while estimated pebble accretion {\em rates} can be very high this is, in part, a consequence of the rapid radial drift of solids which cannot be sustained indefinitely. Which process dominates in the formation of observed planetary systems can only be considered in the context of a global theoretical model for planet formation \citep[e.g.][]{bitsch15,bitsch18}, or by appealing to observational cosmochemical constraints that are only available for the Solar System. \cite{johansen21} argue for a large pebble contribution in terrestrial planet formation, but the division between the two processes in a hybrid Solar System model remains the subject of vigorous debate.

Accretion of the primary atmospheres or envelopes of massive planets occurs (in the consensus view) via the process of {\em core accretion} \citep{mizuno81,ikoma25}. Core accretion is a  broad umbrella that covers the classical planetesimal-only models for Jupiter and Saturn's formation \citep{pollack96}, along with more recent models that can accommodate the existence of the dilute core inferred for Jupiter from JUNO data  \citep{wahl17,helled22}, or a primary role for pebble accretion \citep{lambrechts14}. The key insight underlying all these flavors of core accretion is the predicted existence of a sharp increase in the gas accretion rate when the envelope mass $M_{\rm env}$ exceeds a critical mass, that is approximately set at $M_{\rm env} \sim M_{\rm core}$ by the onset of self-gravity in the envelope \citep{rafikov06}. Below the critical mass, low mass gaseous protoplanets grow relatively slowly, at a rate that is determined by the growth of the core through pebble and planetesimal accretion, and by the time scale on which the envelope can lose thermal energy and contract \citep{ikoma00}. If the gas disk dissipates during this initial phase of core accretion, the expected outcome is a planet that may resemble the Solar System's ice giants, or extrasolar sub-Neptunes. If, instead, the critical mass can be exceeded during the disk lifetime, the planet enters a phase of accelerated gas accretion (this is known as ``runaway growth", although it is still a quasi-hydrostatic evolution) and the outcome is a gas giant with $M_{\rm env} \gg M_{\rm core}$. The minimum core mass that can lead to gas giant formation is primarily a function of the opacity in the envelope, and is roughly $5-10 \ M_\oplus$ \citep{piso15} for grain size distributions that arise in coagulation-fragmentation equilibrium calculations. Smaller critical core values are possible if envelopes are effectively grain-free \citep{hori10}. In the other direction, substantially more massive cores can avoid entering runaway given sufficient heating from planetesimal accretion.

{\em Recycling} of gas between the protoplanetary disk and the outer envelope of a growing planet affects the thermodynamics of the envelope, and may therefore alter the time scale for achieving runaway \citep{ormel15}. The importance of this process depends upon the depth the recycled gas is able to reach within the envelope \citep{savignac24}. Accurate modeling of recycling is a substantial challenge: the flows themselves are intrinsically three-dimensional and can only be captured fully in radiation hydrodynamic simulations, while their impact on core accretion is felt only over long time scales that can only be followed in one dimensional models. Current work suggests that the effect of recycling on the formation of Jupiter itself may be minimal \citep{zhu21b}, but that for close-in (0.1~au) cores there could be as much as an order of magnitude delay in the onset of runaway growth \citep{bailey24}. 

\begin{figure}[t]
\includegraphics[width=\textwidth]{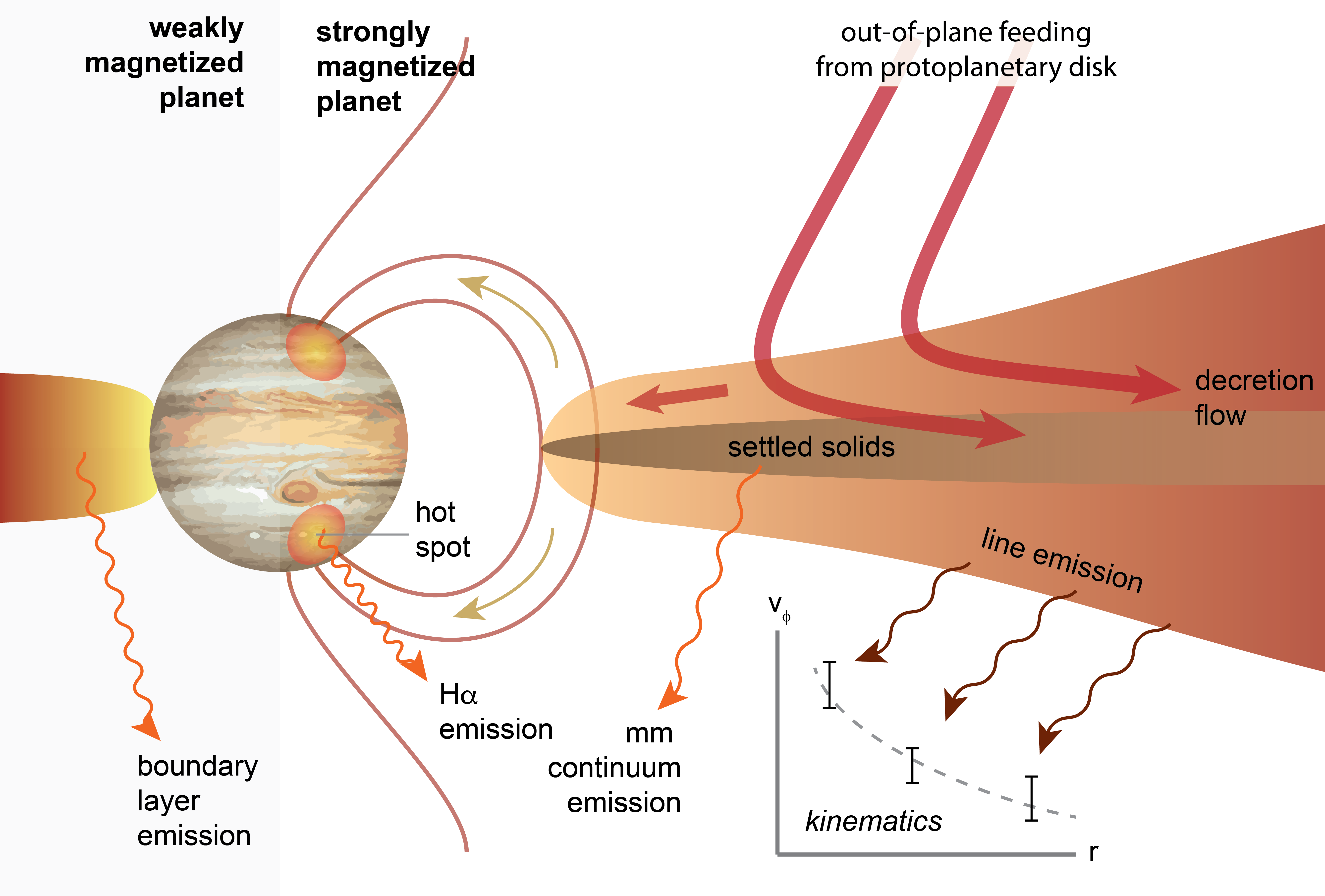}
\caption{Illustration of a possible structure for a circumplanetary disk surrounding a massive planet. Gas is supplied to circumplanetary disks through out-of-plane flows that circularize at a small fraction of the planet's Hill sphere. The inner disk accretes, while the outer disk may be approximately static (if it is tidally truncated) or decreting (if tidal truncation is inefficient). The disk close to the planetary surface terminates in an equatorial boundary layer if the planet is weakly magnetized, while strongly magnetized planets accrete via a magnetosphere that is analogous to accretion in T~Tauri stars. Circumplanetary disks may be detected via dust continuum or molecular line emission, or via accretion signatures.}
\label{Armitage_cpd_figure}
\end{figure}

Following runaway, growth of giant planets to Jovian or super-Jovian masses is ultimately limited by the ability of the planet to accrete from across the gap that opens within the protoplanetary disk. Circumplanetary disks, fed by out-of-plane flows from the protoplanetary disk \citep{tanigawa12}, may form during this final phase of accretion. Circumplanetary disk formation requires not only a sufficiently large planet mass relative to the thermal mass \citep[roughly, a mass such that $r_{\rm Hill} > h$; e.g.][]{sagynbayeva24}, but also favorable thermodynamic conditions that allow fast enough cooling on sub-Hill scales \citep{ayliffe19,szulagyi16,fung19}. Multi-fluid radiation hydrodynamic simulations suggest that cooling on a time scale that is an order of magnitude shorter than the orbital time scale is necessary for gas to form a near-Keplerian disk on scales of $r_{\rm Hill} / 3$ \citep{krapp24}. Observationally, the detection and characterization of circumplanetary disks is in its infancy. PDS~70c, a young massive exoplanet in a system whose protoplanetary disk has a transition disk morphology \citep{keppler18}, shows sub-mm emission at the location of the planet that is consistent with circumplanetary disk expectations \citep{isella19,benisty21}. H$\alpha$ emission is detected from both PDS~70b and PDS~70c \citep{haffert19}, confirming that these are massive planets with ongoing accretion onto the planetary surface. As illustrated in Figure~\ref{Armitage_cpd_figure}, continuum emission is just one of many potential tracers of circumplanetary disks, which may also be observable in molecular  (kinematic) or atomic line emission, or even through the production of jets \citep{gressel13,zhu15}.

The consensus view is that the Solar System's planets, and at least the vast majority of extrasolar planets, formed via core accretion (which, as noted already, is today more of a broad guideline for planet formation rather than a precisely specified rule). It is possible that some giant planets, especially those approaching the $13 \ M_J$ threshold that conventionally demarcates planets from brown dwarfs, instead formed via disk instability \citep{boss97}. The formation of protostellar disks may well lead to initially self-gravitating disks \citep{lin90,xu21}, which have been inferred observationally from kinematic data \citep{speedie24} and seen to fragment, at least into multiple {\em stellar} systems \citep{tobin16}. The physics of self-gravitating disks is reviewed by \cite{kratter16}. Massive disks fragment if the cooling time scale is sufficiently short \citep[of the order of the orbital time scale;][]{gammie01,rice05}, conditions that occur only at large ($\sim 50$~au) radii \citep{clarke09,rafikov09}. What are currently the most realistic radiation hydrodynamics simulations, by \cite{xu24}, find that the distribution of {\em initial} fragment masses in this regime peaks at about $20 \ M_J$. Further accretion in what remains a gas-rich environment post-fragmentation would likely increase fragment masses further \citep{kratter10}. \citep[This fate could only be averted if fragments migrate inward rapidly and are partially tidally stripped;][]{nayakshin10,baruteau11,zhu12b}. Protostellar disks can also be driven to fragment by sufficiently rapid infall \citep{kratter10b}, but this also leads to characteristic masses that are typically in the brown dwarf or low-mass star regime.

\section{Stability and planetary system evolution}
Prescient work by \cite{goldreich80} and \cite{fernandez84} identified physical processes that can lead to the large-scale orbital migration of planets that interact gravitationally with gas or planetesimal disks. Their ideas, along with tidal and purely dynamical processes that can also reshape planetary orbits, received renewed interest with the discovery of the first exoplanets \citep{mayor95}, whose close-in and (separately) often eccentric orbits are inconsistent with our accepted understanding of in situ formation.

Planet-disk interactions, reviewed recently by \cite{paardekooper20}, are one broad class of processes that lead to planetary system evolution. The most intensively studied case involves resonant interactions that allow the exchange of angular momentum with the protoplanetary disk, and migration of the planetary orbit. There are two sub-cases:
\begin{itemize}
 \item Gas at interior and exterior {\bf Lindblad} resonances is excited by the planetary potential \citep{goldreich79}, forming density waves that manifest as one-armed spirals in the disk \citep{ogilvie02}. It is by now well-understood. Key features of the Lindblad torque are that its strength is largely independent of disk structure \citep[in the sense that how the waves dissipate is immaterial;][]{meyer-vernet87}, and that the radial pressure gradient generates a generic asymmetry between the summed inner and outer torques \citep{ward97}. These properties mean that -- although {\em other} torques have more complex behavior -- the total torque on a planet from its interaction with a gaseous disk can only vanish coincidentally, or at special locations, known as migration traps \citep{hasegawa11}. At planet masses low enough that the perturbation to the disk surface density profile is small (the ``Type I" regime), the Lindblad-only contribution to migration is almost always inward.
 \item Gas in the {\bf co-orbital} region exchanges angular momentum with the planet as it executes horseshoe turns in the frame co-rotating with the planet. Asymmetries in the behavior of the turns in the regions that lead / trail the planet, due to radial gradients in entropy and vortensity, lead to a net torque. Complexity arises because, in the absence of viscosity or wind-driven radial motion, the streamlines in the horseshoe region are closed, and would rapidly adjust to null out (``saturate") the torque. Simulations show that adding the co-orbital torque to the Lindblad torque can lead to outward migration, but that this result depends not just on the disk radial structure, but also on how energy and angular momentum are transported \citep{paardekooper10,paardekooper11}.
\end{itemize}
More recently, significant non-resonant sources of migration torque have been identified. A luminous planet generates perturbations in temperature and density in its vicinity, that are asymmetric on either side of the orbit because (yet again) of the radial pressure gradient. This leads to a {\em thermal torque}, which has an established linear theory \citep{masset17}. Besides altering the migration rate \citep[in the sense of modestly broadening the range of parameters for which outward migration occurs;][]{guilera19,baumann20}, an interesting property of thermal torques is their propensity to excite eccentricity and inclination \citep{fromenteau19,cornejo23}. Another relatively new area of investigation is the role of {\em pebble torques}, that arise from the scattering of aerodynamically coupled particles that drift past a planet \citep{benitez-llambay18,chrenko24}. These can also give rise to enhanced outward migration.

The coupled response of the planet-disk system to migration torques has traditionally been studied in the limit where angular momentum transport in the disk is due to turbulence, and the strength of that turbulence is moderately high ($\alpha \sim 10^{-3} - 10^{-2}$). In this case, migration occurs in two regimes. Low mass planets do not open a gap, and migrate in the ``Type I" regime due to the intrinsic asymmetry between inner and outer Lindblad resonance locations \citep{ward97}. High mass planets open a gap, and their migration in the ``Type II" regime is then coupled to the evolution of the disk \citep{lin86}. Whether a planet opens a gap, and hence what regime of migration applies, depends on a simple function of the planet-to-star mass ratio, disk scale height, and $\alpha$-parameter \citep{kanagawa15}. The accuracy of simplified (one dimensional) time-dependent models for Type II migration remains debated \citep{duffell14,scardoni20,kanagawa20}, but it is established that the fastest migration typically occurs in the Type~I regime just prior to gap opening \citep{ward97,bate03}. In many planet formation applications, the inferred migration time scale is comparable to or shorter than the gas disk lifetime.

In low-viscosity disks, the picture is more complex. Planet-disk interaction in this regime commonly forms vortices through the Rossby Wave Instability \citep{lovelace99}, which can themselves migrate and interact with the planet. \cite{mcnally19} show how the migration regime varies in a two-dimensional plane, with the planet mass and disk viscosity as the controlling parameters. Particularly interesting is the case of a disk with a low turbulent viscosity, but efficient wind angular momentum loss. Wind angular momentum loss is essentially a surface rather than a volume process, so a wind-driven disk is expected to drive a rapid radial flow through planetary gaps that will affect migration torques \citep{mcnally20,kimmig20,speedie22,wu25,hu24}. Using three dimensional non-ideal magnetohydrodynamic simulations, \cite{wafflardfernandez23} find that magnetic flux concentration in the gap region leads to an asymmetric gap profile that leads to outward migration of massive planets.

In the linear, Type~I regime, the disk torque on low mass planets is expected to be largely independent of the multiplicity of a planetary system. The migration of individual planets in a system will, however, be affected if there is angular momentum exchange due to mean-motion resonances between the planets. {\em Resonant capture} \citep{goldreich66,borderies84} occurs when there is convergent migration into a mean motion resonance, with a capture probability that depends on the migration speed, eccentricity, and strength of perturbations (due, in the case of migrating planets, to gravitational coupling to disk turbulence) \citep{quillen06,adams08,batygin15}. For giant planets in the Type~II regime, an additional effect arises because overlap of the gaps created by the planets means that the Lindblad torques acting on them are no longer independent. Under the right conditions, two giant planets, that individually would each migrate inward, migrate outward as a coupled resonant pair \citep{masset01}. These rather subtle multi-planet effects are of central importance to the ``breaking the chains" model for close-in Kepler systems \citep{izidoro17}, for modern version of the Nice model for the outer Solar System \citep{tsiganis05,morbidelli07}, and in the Grand~Tack model for the inner Solar System \citep{walsh11}. Planetesimal scattering, subsequent to gas disk dispersal, provides a mechanism for breaking resonances, especially in the context of the Nice model. Given a sufficiently massive planetesimal disk, scattering may be important across a broader range of radii during planet formation \citep{minton14,jinno24}.

Migration is a phenomenon that, once protoplanets exceed masses of a fraction of an Earth mass, impacts all phases of planet formation and planetary system evolution. It has a similar relation to planet formation as opacity does to stellar structure, in that it is critically important, complex, but in principle calculable from first principles to useful precision. There are only a few circumstances where completely ignoring migration in planet formation is a defensible position.

After the dissipation of the protoplanetary disk, planets with atmospheres or envelopes can lose mass through photoevaporation \citep{owen13} and / or core-powered mass loss \citep{ginzburg16}. The immediate architecture emerging from the disk phase -- during which planet-disk interactions typically (though not always) damp eccentricity and inclination -- may also prove to be unstable over long time scales. The simplest model problem here is the evolution of a system of $N$ planets, initially on circular, coplanar orbits, under Newtonian point-mass gravity. For $N=2$, stability to close encounters between the planets (``Hill stability") can be proven beyond a threshold separation \citep{gladman93}. For $N \ge 3$, model planetary systems with equal mass planets, uniformly spaced in $\log(a)$, are found through N-body integrations to be unstable, with a time scale to instability that is a steep function of the initial separation \citep{chambers96,smith09,obertas17}. Although a variety of empirical functions have been proposed, recent results suggest that the instability time is best fit by a power-law function of separation, where the separation is measured in units of the planet mass to the 1/4 power \citep{lammers24}. These numerical findings are interpreted within a dynamical framework of chaos driven by the overlap of resonances \citep{lecar01,quillen11,yalinewich20,petit20,lammers24}. Resonant chain systems -- arguably the most natural initial condition emerging from the disk phase -- have much longer lifetimes than similar planetary systems with non-resonant period ratios. \cite{pichierri20} and \cite{goldberg22} analyze the stability of this important special case.

The {\em outcome} of dynamical instability in a planetary system can be ejection of one or more of the planets, physical collisions, or excitation to a high eccentricity state that may collide or interact tidally with the host star. The parameter that controls the branching ratios between these outcomes is the ratio of the orbital velocity to the escape speed from the surface of the planet. High values of $v_{\rm K} / v_{\rm esc}$ (low mass planets, and / or small orbital radii) favor collisions, while low values lead to a predominance of ejections \citep{ford01,petrovich14}. Collisions dominate for systems of close-in super-Earth and sub-Neptune mass planets, whose properties resemble the outcomes seen in simulations of richer, initially unstable, planetary systems \citep{pu15}.

Planets that are sufficiently widely separated to be quasi-stable to scattering can still exhibit non-trivial secular evolution. Secular dynamics approximates conserves the angular momentum deficit \citep[for a modern reference, see e.g.][]{laskar97},
\begin{equation}
    {\rm AMD} \equiv \sum_k \Lambda_k 
    \left( 1 - \cos i_k \sqrt{1 - e_k^2} \right),
\end{equation}
where the sum is over planets with angular momenta $\Lambda_k$, semi-major axes $a_k$, eccentricities $e_k$, and inclinations (relative to the invariable plane) $i_k$. AMD conservation ensures that no interesting dynamics develops from initially near-circular, near-coplanar systems of widely separated planets \citep[``AMD stability";][]{laskar17}. Given non-zero AMD, however (potentially arising from prior planet-planet scattering events), chaos driven by secular resonances leads to each planet eventually experiencing substantial excursions in AMD about an equipartition value \citep{wu11}. A large AMD for an inner, relatively low mass planet, can lead to a high eccentricity that places the pericenter distance close enough to the star for tidal dissipation to become important. Qualitatively similar behavior is possible under Kozai-Lidov dynamics \citep{kozai62,lidov62,naoz11}, in which a planet exhibits cyclic variations in eccentricity due to secular perturbations from a distant, inclined companion. These secular processes can naturally produce high (and even retrograde) inclinations as well as high eccentricities \citep{wu03,fabrycky07,naoz11}, making them prime suspects in the formation of the misaligned exoplanets that have been detected via the Rossiter-Mclaughlin effect \citep{winn10}.

Dynamical processes are generally considered to play a dominant role in the excitation of giant planets to high eccentricities \citep{rasio96}, and a substantial / dominant role (when combined with the action of tides) in the formation of hot Jupiters \citep{dawson18}. With the detection of planets around younger stars, and better age dating of older planetary systems, there is an exciting possibility to measure and interpret dynamically differences in planetary system architecture as a function of stellar age \citep{hamer24,dai24}. Nonetheless, thirty years on from the discovery of the first exoplanets, the relative importance of the many processes that could lead to hot Jupiter formation remains an open problem.

\section{Some open questions}
The other contributions to this section of the {\em Handbook of Exoplanets} review planet formation theory in much more detail than is possible here. In many areas there has been substantial recent progress. Theoretical advances, many inspired by problems thrown up by new Solar System, disk, and exoplanet observations, have led to a theory of planet formation that inherits many features of classical models while being more efficient and accommodating the formation of diverse final architectures. It is self-evident, however, that much remains to be understood. Open questions and problems that I would like to see resolved include, in no particular order:
\begin{enumerate}
    \item {\em What sets the balance between pebbles and planetesimals?} Aerodynamically coupled particles, especially those with $\tau_{\rm S} \sim 0.1$, are valuable currency in contemporary theory, both for planetesimal formation via the streaming instability and for pebble accretion. Pebbles can only be spent once, however, and it is not clear how the balance is determined.
    \item {\em Uncertainty quantification of planet migration rates.} Computing the rate of gas disk migration to a reliably known level of precision, across the range of planet masses and disk structures relevant to planet formation, would remove a major source of uncertainty in planet formation theory. Code comparison efforts are a valuable step toward this goal.
    \item {\em Is planet formation smoothly distributed or radially concentrated at special disk locations?} Protoplanetary disks contain ice lines, and probably other deterministic radial locations where the disk physics creates particle traps or planet migration null points. How important these ``special" locations are remains rather unclear. Put simply, the zeroth order paradigm for planet formation could yet be either smooth or ring-like.
    \item {\em Do disks evolve mostly due to turbulent or wind torques?} Magnetized winds are currently favored, but even if the basic question is considered resolved, the strength and nature of the turbulence in {\em primarily wind-driven disks} is still of fundamental importance. It matters, among other things, for planetesimal formation, pebble accretion, and planet migration.
    \item {\em Does planet formation overlap spatially or temporally with disk accretion outbursts, or with regions of active gravitational instability?} Either of these processes -- conventionally but not necessarily occurring mostly at early times -- lead to a fundamentally different environment for growth.
    \item {\em What is the structure and evolution of gas and solids in circumplanetary disks?} This is a physics-rich problem with a wide range of relevant spatial scales. Informative constraints from circumplanetary disk, and possibly exomoon observations, will be needed to inform theoretical work. 
    \item {\em Does the composition of the upper levels of exoplanet atmospheres retain formation information?} The division of abundant elements such as carbon and oxygen between condensed and gaseous phases is expected to vary with temperature, and thus radius, in the disk. This opens up the possibility of using atmospheric composition measurements from transmission spectroscopy as a new probe of planet formation \citep{oberg11}. Multiple confounding factors, in both planet formation and atmospheric evolution, might however frustrate such hopes. The Chapter by Pudritz, Cridland, and Inglis reviews this question in depth.
    \item {\em How can theoretical models of planetary system evolution be tested against population-level data?} At the most ambitious level, this requires {\em population synthesis}, reviewed in this volume by Burn and Mordasini, in which all sources of theoretical and observational uncertainty must be considered. However, even in more limited comparisons when the theoretical model is well-specified and accurately calculable (e.g. pure N-body dynamics), and the data well-understood, inferring principled statistical constraints is often difficult. New statistical and Machine Learning approaches show promise.
    \item {\em What combination of processes lead to planetesimal formation?} Is the streaming instability able to form planetesimals -- across the range of disk radii and time necessary for planet formation -- on its own, or are other processes (such as particle concentration in large-scale vortices) that ``pre-concentrate" solids necessary?
    \item {\em What are the properties of planetesimals that form from the streaming instability, and from alternative planetesimal formation channels?} Important quantities such as the initial mass function, and the distribution of binary inclinations, have already been derived from simulations of the streaming instability. How robust these predictions are, and to what extent they discriminate between models, is not completely clear: few simulations include collisional physics or have the resolution to follow collapse to anything close to material density.
    \item {\em How is the opacity, especially that from grains, determined within planetary envelopes?} The efficiency of cooling, determined largely by the envelope opacity, controls the outcome of core accretion, and hence the predicted populations of sub-Neptunes, ice, and gas giants. 
    \item {\em Are the conclusions of N-body, and joint statistical / N-body simulations of planet formation, robust?} GPU-enabled N-body codes can evolve fully self-gravitating systems with substantially larger $N$ than was previously possible, allowing both a test of the robustness of existing results and the possibility of direct simulations of growth across a broader mass range.
    \item {\em Is the material physics of collisions well enough understood for planet formation applications?} The experimental determinations of the threshold velocity for fragmentation of silicates and water ice, and more broadly the outcomes of different collisions between aggregates of various sizes, are critical inputs for planet formation theory. Any revision to the currently accepted paradigm, perhaps from experiments with more realistic materials, could have a major impact. 
\end{enumerate}

\begin{acknowledgement}
My understanding of planet formation has been greatly informed by the work and ideas of students, postdocs, and other collaborators over many years. Thanks to all of you! My work has been supported by NASA, the NSF, and the Simons Foundation. I thank Ralph Pudritz for a careful review of this Chapter.
\end{acknowledgement}

\bibliographystyle{spbasicHBexo}  

\end{document}